\documentclass[epj]{svjour}
\usepackage{amsmath}
\usepackage{amssymb}
\usepackage{amsfonts}
\usepackage{graphics}
\begin{document}
\title{Field emission with relativistic effects in a magnetic field}
\author{S. Lebedynskyi \and R. Kholodov% etc
}                     % Do not remove
\institute{Institute of Applied Physics, National Academy of Sciences of Ukraine, Petropavlivska str. 58, 40000 Sumy, Ukraine}
\date{Received: date / Revised version: date}
\abstract{
The work concerns relativistic effects and the influence of an external magnetic field on the transmission coefficient. The Fowler-Nordheim equation has been relativistically generalized and effect of the Lorentz contraction of a potential barrier at the metal-vacuum interface has been found. Influence of the magnetic field parallel to a metal surface on the transmission coefficient is taken into account when $cB < E$.
 % end of PACS codes
} %end of abstract
\maketitle
\section{Introduction}
\label{intro}

Modern experiments in elementary particle physics require high energies. It results in construction of accelerators with high accelerating ingredients  \cite{Accomando98}. Experiments with the accelerating structures of the compact linear collider (CLIC) prototype have shown a breakdown occurrence on these gradients testing \cite{Kidemo}. Field electron emission is believed to play the main part at the first stage of the breakdown formation \cite{Latham,Ilic,Djurabekova17,Djurabekova18}. Theory of the field electron emission based on a quantum-mechanical tunneling of electrons through a surface potential barrier was developed by R.~H.~Fowler and L.~W.~Nordheim in 1928 \cite{Fowler} but it still remains the main theory used to calculate field emission current density for the time being. Electron motion becomes relativistic as voltage and interelectrode gaps increase. The relativistic effects should be considered to develop a general approach describing tunneling of the electrons through the potential barrier and to refine an expression for a transmission coefficient for tasks with high fields and work function (for example, electron emission from the polar region of strongly magnetized neutron stars).

Various options for increasing the breakdown resistance in the accelerator's structural materials were considered in literature. Metal surface modification \cite{Baturin}, surface conditioning \cite{Descoeudres}, and application of a magnetic field \cite{Lebedynskyi18} are among them. This paper concerns suppression of field electron emission current by an external magnetic field which is parallel to a metal surface.

Influence of the magnetic field on the emission current has been previously theoretically studied in \cite{Blatt,Gogadze,Buribaev,Lebedynskyi2,Lebedynskyi3,Ghosh2}. Blatt~F.J. \cite{Blatt} considered field emission from a plain surface of a metal in an external magnetic field perpendicular to it. Blatt assumed that transparency of the potential barrier on the metal-vacuum interface is independent from the magnetic field. This assumption was proved in \cite{Lebedynskyi2}. Field emission current decreases as $B^2$ in compliance with an expression for current density given in \cite{Gogadze}. This decrease is also complemented by periodic fluctuations in current. The oscillations of the field emission current in a magnetic field which is perpendicular to the metal surface were theoretically studied in \cite{Buribaev}. 
In one of the first works on the modification of Fowler-Nordheim cold emission in presence of strong magnetic field \cite{Ghosh2}, the authors have studied the emission of electrons from the polar region of strongly magnetized neutron stars. 
However, influence of the magnetic field parallel to the metal surface has not been theoretically studied yet.

This work derives a relativistic expression for a transmission coefficient for the electron tunneling through the potential barrier, with influence of a parallel magnetic field considered. Wave functions of an electron motion at relativistic speed to be found, the Klein-Gordon equation instead of the Schrödinger equation was used. The Klein-Gordon equation allows consideration of the external uniform magnetic field parallel to the metal surface but neglects the electron spin. The case when the charged particle just deflects under the influence of the magnetic field and moves to infinity under the action of an electric field (that is when ${{E}^{2}}-{{(cB)}^{2}}>0$) is considered.

%==============================================================================

\section{Relativistic generalization of the transmission coefficient}

\label{sec:Rel}
To describe the electron tunneling we choose a Cartesian system with the electric field strength vector $\vec{E}(-E,0,0)$ directed along an $x$-axis. The transmission coefficient of the potential barrier at the metal-vacuum boundary ($x=0$) can be written as:
\begin{equation}
\label{DW}
D=\frac{{\left| A \right|^2}-{\left| B \right|^2}}{{\left| A \right|^2}}, 
\end{equation}
where $A$ and $B$ are the amplitudes of the incident and reflected waves in metal $(x<0)$. We can find them from the wave function of electron, which has a known form \cite{Akhiezer}:
\begin{equation}
\label{psi1}
\psi_1(x) =A  \exp{\left(\frac{i}{\hbar}p_1x\right)}+B  \exp{\left(-\frac{i}{\hbar}p_1x\right)},
\end{equation}
where ${{p}_{1}}$ is the component of the electron momentum along an electric field.

The wave function of the electron in vacuum $(x>0)$ in the presence of an electric field satisfies the Klein-Gordon equation \cite{Akhiezer}:
\begin{equation}
\label{KG}
\left[ \frac{\partial^2 }{\partial \xi^2}+ \frac{\xi ^2}{4} -a\right] \psi_2(\xi)=0 ,
\end{equation}
where $ \frac{\xi}{\sqrt{2}}=\sqrt{\frac{|e|E}{\hbar c}} \left(x+\frac{\varepsilon -U_0}{|e|E} \right)$, $\varepsilon=mc^2+W_e$, $a=\frac{m^2c^3}{2\hbar e E}$, ${{U}_{0}}$ is the potential barrier height, $E$ is electric field strength, $W_e$ is the electron’s kinetic energy.

Equation (\ref{KG}) is a parabolic cylinder equation \cite{Watson}. We seek the solution of equation (\ref{KG}) which represents a wave traveling to the right at large $x$
\begin{equation}
\label{psi2}
\psi_2(\xi)=D_{-1/2-ia}\left(\xi e^{-\frac{\pi i}{4}} \right) .
\end{equation}

Wave function matching at the metal-vacuum boundary is used to find the transmission coefficient (See Appendix). The transmission coefficient of the potential barrier (\ref{DW}) takes the form:
\begin{equation}
\label{Drel}
D_{rel}=\frac{2\sqrt{2}\beta\frac{p_1}{\hbar}\left(\frac{\hbar c}{eE} \right)^{1/2} }{\alpha^2+\beta^2+\frac{p_1^2}{2\hbar^2}\frac{\hbar c}{eE}+\sqrt{2}\beta\frac{p_1}{\hbar}\left(\frac{\hbar c}{eE} \right) ^{1/2}},
\end{equation}
where $\alpha$, $\beta$ are real and $\alpha+i\beta=\frac{dD_{-1/2-ia}\left[e^{-\frac{\pi i}{4}} Q \right]}{dQ}$.

The transmission coefficient of the potential barrier (\ref{Drel}) can be found only numerically in the general case. The calculations have been carried out according to the general expression of the transmission coefficient without additional assumptions. We use the following parameter values hereafter:
\begin{eqnarray}
\label{FE1}
E\cong 10^9 \frac{V}{m}, \phi \cong 5 eV,
\\
\label{FE2}
E\cong 10^{16} \frac{V}{m}, \phi \cong 100 keV.
\end{eqnarray}

Here, (\ref{FE1}) are the typical values for laboratory conditions, (\ref{FE2}) are typical values for field emission from neutron stars surface \cite{Beskin,Diver,Ghosh1}.

It is easy to see that in the case of the field emission from metals (\ref{FE1}),the magnitude of the relativistic correction is less than $0.1\%$. Of course, the experimental observation of such effect is not feasible, taking into account exponential increase in the field emission current density. In the case of strong field (\ref{FE2}) approaching the Schwinger limit ${{E}_{s}}\simeq 1.32\times {{10}^{18}}~ {V}/{m}$ and work function of $100\:keV$, the difference in transmission coefficient exceeds $10\%$ and can make a significant contribution to the current density.

\begin{figure}
	\resizebox{\columnwidth}{!}{\includegraphics{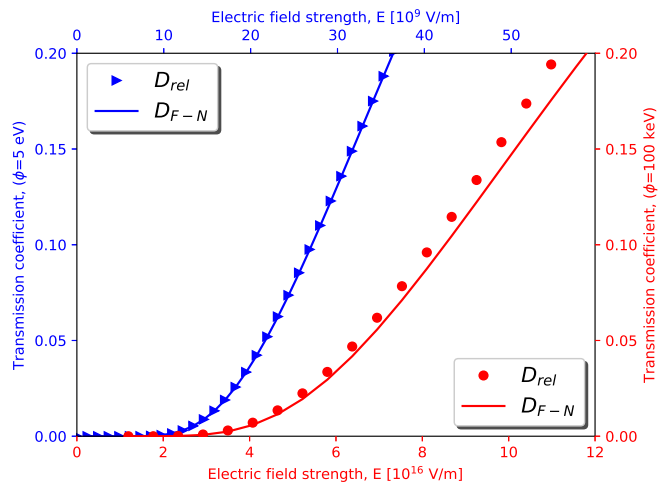}}
	\caption{%
		Numerical calculation of the obtained by Fowler-Nordheim and relativistic generalized transmission coefficients for different work function values.
	}
	\label{fig:Drel}
\end{figure}

It is possible to find a simple analytical expression for the formula (\ref{Drel}) when
\begin{equation}
\label{muL}
\left\lbrace
\begin{split}
\frac{\phi}{mc^2}\ll 1 \\
\frac{\phi^2}{eE\hbar c} \sim 1
\end{split}
\right. .
\end{equation}

It should be pointed out that both conditions (\ref{FE1}) and (\ref{FE2}) satisfies (\ref{muL}).The transmission coefficient of a potential barrier at the metal–vacuum boundary is accurately up to second-order terms (See Appendix) can be written as
\begin{equation}
\label{DREL}
\begin{split}
{{D}_{rel}}={{D}_{F-N}}\left( 1+\frac{\sqrt{2}}{5}\frac{{{\left( U_0-W_e \right)}^{\frac{5}{2}}}}{\sqrt{m}{{c}^{2}}eE\hbar } \right. \\ +\frac{\sqrt{2}}{48}\frac{\left( 7U_0-12W_e \right)eE\hbar }{U_0\sqrt{m}{{\left( U_0-W_e \right)}^{{}^{3}/{}_{2}}}}+
\frac{1}{120}\frac{37U_0-79W_e+12\frac{{{W_e}^{2}}}{U_0}}{m{{c}^{2}}}
\\
+\left.
\frac{1}{1536}\frac{\left( 49{{U_0}^{2}}-216U_0W_e+192{{W_e}^{2}} \right){{e}^{2}}{{E}^{2}}{{\hbar }^{2}}}{{{U_0}^{2}}m{{\left( U_0-W_e \right)}^{3}}}
\right),
\end{split}
\end{equation}
\begin{equation}
\label{DFN}
{{D}_{F-N}}={4\sqrt{U_0-W_e}\sqrt{W_e}{U_0}^{-1}{{e}^{-\frac{4}{3}\frac{{{(U_0-W_e)}^{\frac{3}{2}}}\sqrt{2m}}{\hbar eE}}}}.
\end{equation}
Equation (\ref{DFN}) is the transmission coefficient obtained by Fowler and Nordheim \cite{Fowler}. At the same time, we note that the second and the last terms of the expression (which do not contain the speed of light c) are the correction to the expression for the transmission coefficient obtained by Fowler and Nordheim and can be obtained from their calculations. The first and the third terms are purely relativistic and cannot be obtained in the framework of the Fowler and Nordheim approach. We also note that the first term is the relativistic correction previously obtained in the case of small interelectrode distances \cite{Lebedynskyi4}.

Figure~\ref{fig:Drel} shows the dependence of the potential barrier transmission coefficient on the electric field strength. The magnitudes of the corrections in the case of field emission from metals (\ref{FE1}) and field emission from neutron stars surface (\ref{FE2}) (conditions (\ref{muL}) are satisfied) are respectively $0.015\%$ and $15 \%$, in accordance with numerical calculations. It can be concluded that in the laboratory conditions the relativistic correction makes an extremely small contribution and is not experimentally observable. But in the case of field emission from neutron stars surface, the contribution is substantial and should be taken into account. Also let's note an increasing of the transmission coefficient due to relativistic corrections obtained in equation (\ref{DREL}).

%===============================================================================
\section{THE INFLUENCE OF A MAGNETIC FIELD ON THE TRANSMISSION COEFFICIENT} \label{sec:MF}

The Klein-Gordon equation for external mutually perpendicular electric $\vec{E} (-E,0,0)$ and magnetic $\vec{B}(0,B,0)$ fields has form (\ref{KG}) with following symbols:
\\ $\frac{\xi }{\sqrt{2}}={{\left( \frac{{{e}^{2}}\left( {{E}^{2}}-{{c}^{2}}{{B}^{2}} \right)}{{{c}^{2}}{{\hbar }^{2}}} \right)}^{\frac{1}{4}}}\left( x-{{x}_{c}} \right),$ $
{{x}_{c}}=-\frac{\left( \varepsilon -{{U}_{0}} \right)E}{e\left( {{E}^{2}}-{{c}^{2}}{{B}^{2}} \right)},$\\ 
$a =\frac{{{\left( \left( \varepsilon -{{U}_{0}} \right)E+B{{}^{2}}{{p}_{2}} \right)}^{2}}}{2ec\hbar {{\left( {{E}^{2}}-{{c}^{2}}{{B}^{2}} \right)}^{\frac{3}{2}}}}-\frac{{{\left( \varepsilon -{{U}_{0}} \right)}^{2}}}{2ec\hbar \sqrt{{{E}^{2}}-{{c}^{2}}{{B}^{2}}}}+\frac{{{m}^{2}}{{c}^{3}}+p_{2}^{2}c}{2e\hbar \sqrt{{{E}^{2}}-{{c}^{2}}{{B}^{2}}}}$ \\ and ${{p}_{2}}$ is the component of the electron momentum along a magnetic field.

The transmission coefficient can be written as: 
\begin{equation}
{{D}_{B}}=\frac{2\sqrt{2}\beta \frac{{{p}_{1}}}{\hbar }{{\left( \frac{{{c}^{2}}{{\hbar }^{2}}}{{{e}^{2}}\left( {{E}^{2}}-{{c}^{2}}{{B}^{2}} \right)} \right)}^{\frac{1}{4}}}}{{{\alpha }^{2}}+{{\beta }^{2}}+\frac{{{p}_{1}}^{2}}{2{{\hbar }^{2}}}\frac{c\hbar }{e\sqrt{\left( {{E}^{2}}-{{c}^{2}}{{B}^{2}} \right)}}+\sqrt{2}\beta \frac{{{p}_{1}}}{\hbar }{{\left( \frac{{{c}^{2}}{{\hbar }^{2}}}{{{e}^{2}}\left( {{E}^{2}}-{{c}^{2}}{{B}^{2}} \right)} \right)}^{\frac{1}{4}}}}
\end{equation}
and it concurs with (\ref{Drel}) in the case $B=0$ and $p_2=0$.

With respect to conditions (\ref{muL}), the transmission coefficient in the first nonvanishing approximation (See Appendix) can be written as
\begin{equation}
\label{DB}
\begin{split}
{{D}_{B}}={{\text{e}}^{-\frac{4}{3}\frac{\sqrt{2}\left( {{E}^{2}}-2\left( \frac{U_0-W_e}{m {{c}^{2}}} \right){{c}^{2}} {{B}^{2}} \right) \sqrt{m}{{\left( U_0-W_e \right)}^{3/2}}}{{{E}^{3}}eh}}} \\
\times \frac{4\sqrt{U_0-W_e}{{\left( {{E}^{2}}-{{B}^{2}}{{c}^{2}} \right)}^{3/4}}{{E}^{3/2}}\sqrt{W_e}}{W_e{{E}^{3}}+{{\left( {{E}^{2}}-{{B}^{2}}{{c}^{2}} \right)}^{3/2}}\left( U_0-W_e \right)}.
\end{split}
\end{equation}

In the case of $B=0$ equation (\ref{DB}) coincides with the transmission coefficient obtained by Fowler and Nordheim \cite{Fowler}. 
\begin{figure}
	\resizebox{\columnwidth}{!}{\includegraphics{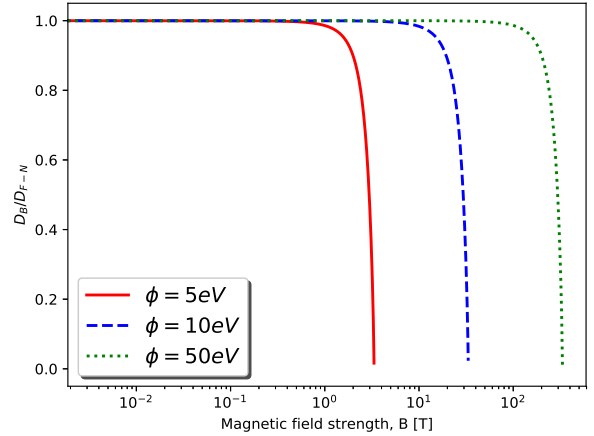}}
	\caption{%
		Dependence of the transmission coefficient upon magnetic field strength for the different work function values.
	}
	\label{fig:DB}
\end{figure}

Figure \ref{fig:DB} shows the dependence of the transmission coefficient at the metal–vacuum interface on the induction of an external magnetic field parallel to the metal surface. We can see an insignificant change in transmission coefficient at small values of magnetic field strength. However, the transmission coefficient decreases when induction increases to the value $B=E/c$.

The experiments on suppression of vacuum breakdowns by a magnetic field have been carried out at CERN and IAP NASU with field values of $E=144~MV/m,~B=0.5~T$ and $E=100~MV/m,~B=0.33~T$ respectively \cite{Lebedynskyi18}. It should be noted that the field enhancement factor is usually of about $30\div 140$ in such experiments \cite{Descoeudres2}. Then we can find from (\ref{DB}) that in the case (\ref{FE1}) the transmission coefficient reduction is less than $0.015\%$, which is consistent with the previously obtained results \cite{Lebedynskyi3,Lebedynskyi18}. The transmission coefficient decreases by $0.1\%$, $5\%$ and $50\%$ when $cB$ is equal to $0.1~ E$, $0.5~ E$ and $0.9~ E$ respectively. To obtain experimentally noticeable reduction of the transmission coefficient of about $10\%$, it is necessary to apply magnetic field of order of $10~ T$.

%===============================================================================
\section{EFFECT OF THE LORENTZ CONTRACTION OF THE POTENTIAL STEP} \label{sec:contraction}

To explain the obtained effect of increase of the transmission coefficient due to relativistic effects let us consider a simple problem of the electron tunneling through a rectangular barrier, with potential defined as

\begin{equation}
\left\{ \begin{matrix}
{{U}_{I}}=0,  \\
{{U}_{II}}={{U}_{0}} , \\
{{U}_{III}}=0  .\\
\end{matrix} \right.
\end{equation}	

For each of three domains the Klein-Gordon equation can be as (\ref{KG}). The solutions of this equation are plan waves
\begin{equation}
\begin{split}
\label{psi3}
{{\psi }_{I}}={{A}_{I}}{{e}^{i\sigma x}}+{{B}_{I}}{{e}^{-i\sigma x}}, \\ 
{{\psi }_{II}}={{A}_{II}}{{e}^{\rho x}}+{{B}_{II}}{{e}^{-\rho x}}, \\ 
{{\psi }_{III}}={{e}^{i\sigma x}}, \\ 
\end{split}
\end{equation}
where 
\begin{equation}
\label{params2}
\left\{ 
\begin{split}
\sigma =\sqrt{\frac{W_e}{{{c}^{2}}{{\hbar }^{2}}}\left( 2m{{c}^{2}}+W_e \right)},  \\
\rho =\sqrt{\frac{W_e-{{U}_{0}}}{{{c}^{2}}{{\hbar }^{2}}}\left( 2m{{c}^{2}}+W_e-{{U}_{0}} \right)}.  \\
\end{split}
\right.
\end{equation}
Equations (\ref{psi3}) concur with nonrelativistic equations \cite{Landau} but with relativistic coefficients (\ref{params2}).  The reflection coefficient $R$ can be expressed via the constants $A_I$, $B_I$ as:
\begin{equation}
R={{\left| \frac{{{B}_{I}}}{{{A}_{I}}} \right|}^{2}}.
\end{equation}

Considering (\ref{psi3}) the reflection coefficient can be written as:
\begin{equation}
\label{R1}
R=\frac{s{{h}^{2}}\left( \rho h \right)}{{{\cos }^{2}}\left( 2y \right)s{{h}^{2}}\left( \rho h \right)+{{\sin }^{2}}(2y)c{{h}^{2}}\left( \rho h \right)},
\end{equation}
where $\cos y=\frac{\sigma }{\sqrt{{{\sigma }^{2}}+{{\rho }^{2}}}},\ \ \sin y=\frac{\rho }{\sqrt{{{\sigma }^{2}}+{{\rho }^{2}}}}$ and $h$ is the potential barrier width.

Note that this expression coincide with nonrelativistic values (see \cite{Landau}) and can be written in most widely known form as:
\begin{equation}
\label{R2}
R=\frac{{{U_0}^{2}}(ch(2\rho h)-1)}{{{U_0}^{2}}ch(2\rho h)-8{{W_e}^{2}}+8W_eU_0-{{U_0}^{2}}}.
\end{equation}

However, the relativistic expression for $\rho $ differs from the non-relativistic case. Indeed, $\rho $ can be rewritten as.
\begin{equation}
\rho ={{\rho }_{nonrel}}\sqrt{1-{{{V}^{2}}}/{{{c}^{2}}}},						\end{equation}
where ${{\rho }_{nonrel}}\equiv \sqrt{\frac{2m}{{{\hbar }^{2}}}\left( U_0-W_e \right)}$ and ${{V}^{2}}=\frac{U_0-W_e}{2m}.$

Then the reflection coefficient (\ref{R2}) can be written as
\begin{equation}
R\left( \rho h \right)=R\left( {{\rho }_{nonrel}}{{h}_{rel}} \right),
\end{equation}
where ${{h}_{rel}}=h\sqrt{1-{{{V}^{2}}}/{{{c}^{2}}}}$.

Thus, the relativistic coefficient is described by the same equation as in non-relativistic case but with decreased barrier width, ${{h}_{rel}}<h$. This effect can be compared to Lorentz contraction of the potential barrier width. As a result, the transmission coefficient accordingly increases, which explains the effect obtained in the first chapter. 

%==============================================================================
\section{Summary}

In present work, the Fowler-Nordheim equation for field emission current density has been generalized to the relativistic case. We presented an approximate formula for transmission coefficient of a potential barrier at the metal-vacuum boundary. The approach is similar to the one developed in \cite{Fowler}. The effect of Lorentz contraction of a potential barrier at the metal-vacuum interface was found. This effect results in increasing of the transmission coefficient by $0.015\%$ for field emission in laboratory conditions and by $~15\%$ for emission from neutron stars surface.

This paper also presents a generalization of the Fowler–Nordheim equation, enabling us to take into account the effect of an external magnetic field parallel to the  metal surface on the transmission coefficient. An expression for the transmission coefficient was found when the condition $cB<E$ is satisfied. For typical experimental values of $E$ and $B$ \cite{Lebedynskyi18}, the effect of the magnetic field on the transmission coefficient was found. In the case of field electron emission from metals (\ref{FE1}) the transmission coefficient decreases by $0.1\%$, $5\%$ and $50\%$ when $cB$ is equal to $0.1~ E$, $0.5~ E$ and $0.9~ E$ respectively. 

%==============================================================================
\section{Appendix}

The conditions of continuity of the wave function and of its derivative in the generalized case have the form:
\begin{equation}
\label{sys1}
\left\lbrace
\begin{split}
\left. \psi_1\right|_{x=0}=A+B  \\
\left. \psi_2\right|_{x=0}=D_{-1/2-ia}\left[e^{-\frac{\pi i}{4}}Q  \right] 
\end{split}   \right. ,
\end{equation}
\begin{equation}
\label{sys2}
\left\lbrace
\begin{split}
\left. \frac{d\psi_1}{dt}\right|_{x=0}=\frac{ip_1}{\hbar}(A-B)  \\
{{\left. \frac{d{{\psi }_{2}}}{dx} \right|}_{x=0}}=\sqrt{2}{{e}^{-i\pi /4}}{{\left( \frac{{{e}^{2}}\left( {{E}^{2}}-{{c}^{2}}{{B}^{2}} \right)}{{{c}^{2}}{{\hbar }^{2}}} \right)}^{\frac{1}{4}}}\\
\times D'_{-1/2-ia}\left[e^{-\frac{\pi i}{4}}Q  \right]
\end{split}   \right. ,
\end{equation}
where we introduce the notation $Q=\frac{\sqrt{2}\left( \varepsilon -{{U}_{0}} \right)E}{\sqrt{c\hbar e}{{\left( {{E}^{2}}-{{c}^{2}}{{B}^{2}} \right)}^{3/4}}}$, so that $Q$ is real. Here and further all equations are valid both for $B=0$ and $B\neq 0$. Then
\begin{equation}
\label{dD}
D'_{-1/2-ia}\left[e^{-\frac{\pi i}{4}} Q \right] =e^{\frac{\pi i}{4}}\frac{dD_{-1/2-ia}\left[e^{-\frac{\pi i}{4}} Q \right]}{dQ}.
\end{equation}

Now equation (\ref{sys1}, \ref{sys2}) can be rewritten as
\begin{equation}
\label{sys3}
\left\lbrace
\begin{split}
A-B=-\sqrt{2}\frac{i\hbar}{{{p}_{1}}}{{\left( \frac{e^2{\left( {{E}^{2}}-{{c}^{2}}{{B}^{2}} \right)}}{c^2\hbar^2}\right)}^{\frac{1}{4}}}\left( \alpha+i\beta\right)  \\
A+B= D_{-1/2-ia}\left[e^{-\frac{\pi i}{4}} Q \right]
\end{split}   \right. ,
\end{equation}
where
$$\alpha+i\beta=\frac{dD_{-1/2-ia}\left[e^{-\frac{\pi i}{4}} Q \right]}{dQ}$$
and $\alpha$, $\beta$ are real. Assuming one direction motion of electron inside metal we set $p{{'}_{2}}=0$. It follows from (\ref{sys1},~\ref{sys2}) that ${{p}_{2}}=0$ in this case.

To determine the coefficients $\alpha$ and $\beta$ we will use the connection formula \cite{Abramowitz}:
\begin{equation}
\label{DtoW}
D_{-1/2-ia}\left[e^{-\frac{\pi i}{4}} Q \right]=C
\left(k^{-1/2}W(a,Q)+ik^{1/2}W(-a,Q) \right),
\end{equation}
where $C=\frac{\sqrt{2}}{2}e^{-\frac{\pi a}{2}-\frac{i}{2}(\frac{pi}{4}+\Phi)}$, $\Phi=\arg{\left( 1/2+ia\right) }$, $k=\sqrt{1-\exp{(2\pi a)}}-\exp{(\pi a)}$ and $W(a,Q)$ is the Weber's parabolic cylinder function, which is real-valued for a real argument.

Then, it is easy to show that 
\begin{eqnarray}
\label{alpha}
\alpha=\frac{\frac{1}{k}W'(a,Q)W(a,Q)+kW'(a,-Q)W(a,-Q)}{\left|k^{-1/2}W(a,Q)+ik^{1/2}W(a,-Q) \right| ^2},
\\
\label{beta}
\beta=\frac{W'(a,-Q)W(a,Q)+W'(a,Q)W(-a,Q)}{\left|k^{-1/2}W(a,Q)+ik^{1/2}W(a,-Q) \right| ^2}.
\end{eqnarray}

The numerator of $\beta$ is the Wronskian of Weber's parabolic cylinder function  W$\left\lbrace W(a,Q), W(a,-Q) \right\rbrace=1 $.

Then $\beta$ can be written as
\begin{equation}
\label{beta12}
\beta=\frac{1}{\left|k^{-1/2}W(a,Q)+ik^{1/2}W(a,-Q) \right| ^2}.
\end{equation}
We will use the asymptotic approximation in this case to find the explicit form of $\alpha$ and $\beta$. Since the Schwinger limit ${{E}_{s}}=\frac{{{m}^{2}}{{c}^{3}}}{e\hbar }\simeq 1.32\times {{10}^{18}}~ {V}/{m}$ is much higher than the laboratory values of the electric field strengths,
$a=\frac{{{E}_{s}}}{2E}$ is sufficiently large and we can write:
$$k=\sqrt{1+\exp (2\pi a)}-\exp (\pi a)\approx 0,$$ 
$$ \frac{1}{k}=\sqrt{1+\exp (2\pi a)}+\exp (\pi a)\approx 2\exp (\pi a).$$

Consequently, the explicit expressions for $\alpha$ and $\beta$ reads
$$\alpha=\frac{W'(a,Q)}{W(a,Q)},~
\beta=\frac{1}{\frac{1}{k}W{{(a,Q)}^{2}}}$$.

To evaluate $W(a,Q)$ we can use asymptotic expression for 
$ a \rightarrow +\infty, -1+\delta \leq \frac{Q}{2\sqrt{a}}\leq 1-\delta $
In this case, the Weber's parabolic cylinder function $W(a,Q)$ and its derivative $W'(a,Q)$ have the following form \cite{Olver}:
\begin{equation}
\label{Webber}
W\left( \frac{1}{2} \mu^2, \mu t \sqrt{2}\right) \sim\frac{l(\mu)e^{\mu^2\eta}}{\sqrt{2}e^{\frac{\pi \mu^2}{4}}\left(1-t^2 \right)^{\frac{1}{4}} }\sum_{s=0}^{\infty}\frac{(-1)^s u_s(t) }{\mu^{2s} \left(1-t^2 \right)^{\frac{3}{2}s} } ,
\end{equation}
\begin{equation}
\label{dWebber}
W'\left( \frac{1}{2} \mu^2, \mu t \sqrt{2}\right) \sim \frac{\mu l(\mu)e^{\mu^2\eta}}{{2}e^{\frac{\pi \mu^2}{4}}}\left(1-t^2 \right)^{\frac{1}{4} }\sum_{s=0}^{\infty}\frac{(-1)^s v_s(t) }{\mu^{2s} \left(1-t^2 \right)^{\frac{3}{2}s} } ,
\end{equation}
where
\begin{equation}
\label{param}
\begin{split}
t=\frac{Q}{\sqrt{2\mu}}, \eta=\int_{t}^{1}{\left(1-t^2 \right)^{\frac{1}{2}}dt}, l(\mu)\sim \frac{2^{\frac{1}{4}}}{\sqrt{\mu}}\sum_{s=0}^{\infty}{\frac{l_s}{\mu^{4s}}},\\ l_0=1,
l_1=-\frac{1}{1152}, u_0(t)=1, u_1(t)=\frac{t^3-6t^2}{24},\\
v_0(t)=1, v_1(t)=\frac{t^3+6t^2}{24}, x_0=\frac{U_0-W_e}{mc^2}, L=\mu^2 x_0^2.
\end{split}
\end{equation}
%
%==================================================================================
%
\section{Authors contributions}
All the authors were involved in the preparation of the manuscript.
All the authors have read and approved the final manuscript.
%
% BibTeX users please use
% \bibliographystyle{}
% \bibliography{}

\end{document}